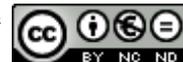

# Simulation of electrochemical processes during oxygen evolution on Pb-MnO₂ composite electrodes

Sönke Schmachtel[a,b,*,1], Lasse Murtomäki[a,**,1], Jari Aromaa[b], Mari Lundström[b], Olof Forsén[b,1], Michael H Barker[c,1]

[a]Department of Chemistry and Materials Science, Aalto University; PO Box 16100, 00076 Aalto, Finland
[b]Department of Chemical and Metallurgical Engineering, Aalto University; PO Box 16200, 00076 Aalto, Finland
[c]VB Consulting Oy, Maaherrankatu 25, Pori 28100, Finland

## *Abstract*

The geometric properties of Pb-MnO₂ composite electrodes are studied, and a general formula is presented for the length of the triple phase boundary (TPB) on two dimensional (2D) composite electrodes using sphere packing and cutting simulations. The difference in the geometrical properties of 2D (or compact) and 3D (or porous) electrodes is discussed. It is found that the length of the TPB is the only reasonable property of a 2D electrode that follows a $1/r$ particle radius relationship. Subsequently, sphere packing cuts are used to derive a statistical electrode surface that is the basis for the earlier proposed simulations of different electrochemical mechanisms. It is shown that two of the proposed mechanisms (conductivity and a two-step-two-material kinetic mechanism) can explain the current increase at Pb-MnO₂ anodes compared to standard lead anodes.

The results show that although MnO₂ has low conductivity, when combined with Pb as the metal matrix, the behaviour of the composite is not purely ohmic but is also affected by activation overpotentials, increasing the current density close to the TPB. Current density is inversely proportional to the radius of the catalyst particles, matching with earlier experimental results. Contrary to earlier SECM experiments, mass transport of sulphuric acid is not likely to have any influence, as confirmed with simulations.

A hypothetical two-step-two-material mechanism with intermediate $H_2O_2$ that reacts on both the Pb matrix and MnO₂ catalyst is studied. It was found that assuming quasi-reversible generation of $H_2O_2$ followed by its chemical decomposition on MnO₂, results are obtained that agree with the experiments. If the quasi-reversible formation of $H_2O_2$ occurs near the peroxide decomposition catalyst, current increases, leading to an active TPB and to the current density that scales with $1/r$. It is further emphasised that both the Pb matrix and MnO₂ catalyst are necessary and their optimum ratio depends on the used current density. Yet, additional experimental evidence is needed to support the postulated mechanism.

* Corresponding author. Tel.: +358 44 020 3377. E-Mail address: soenkes@gmx.de
** Corresponding author. Tel.: +358 50 570 6352. E-Mail address: lasse.murtomaki@aalto.fi
1 ISE member

## *Highlights:*

- The length of the triple phase boundary as a function of the particle size, $r_p$ and the surface coverage, $\Theta$ on 2D composite electrodes ($L \propto \Theta/r_p$).

- Diffusion domain approach extended to randomised composites and to secondary current distribution.

- Use of electrocatalysts with low conductivity in composite anodes.

- A hypothetical two-step-two-material kinetic mechanism leading to lower activation overpotentials during oxygen evolution on composites with $H_2O_2$ as the intermediate.

## *1. Introduction*

During metal electrowinning (EW) from sulphate based electrolytes the oxygen evolution reaction (OER) occurs on the anode. The major metals that are electrowon from sulphate media are zinc, copper, nickel and cobalt [1].

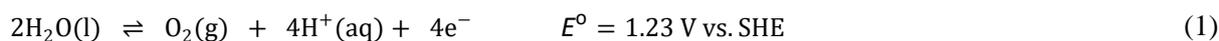

$$2H_2O(l) \;\rightleftharpoons\; O_2(g) \;+\; 4H^+(aq) +\; 4e^- \qquad E^O = 1.23 \text{ V vs. SHE} \qquad (1)$$

In metal EW, when using traditional lead based anodes, the overpotential of the OER is in the range of 500...800 mV [2,3]. It contributes substantially to the cell voltage and to the operating costs. Decreasing the OER overpotential has the advantage that other operational parameters need only minor changes and that anodes are not exposed to highly oxidizing potentials that cause their corrosion [4].

The overpotential of the OER can be lowered using composite anodes [5]. This is especially interesting for Zn EW where lead anodes alloyed with silver are traditionally used, each anode containing ca. 0.5 kg of silver. Alloying lead with other less expensive metals or $MnO_2$ would significantly lower the cost of the anodes. Also, silver has been identified as a critical metal whose demand will exceed its production in the future, which is also seen critical for sustainable energy production [6,7]. Studies on Pb-$MnO_2$ composite materials have addressed the activity of the coating [4,8–19], corrosion rates [4,15,20], kinetics [4,12], and manganese reactions [4,16,19]. From the point of view of manufacturing, alternatives to co-deposition and mixed powder pressing of Pb-$MnO_2$ composites have been considered, including cold sprayed and high velocity oxygen fuel sprayed lead anodes [18]. Some recent studies also included the manufacturing of Pb-$MnO_2$ composite anodes by accumulative roll bonding [10,13]. Other technologies are covered in reference [17].

In our previous paper [17], pressed Pb-$MnO_2$ composites were studied using different $MnO_2$ types ($\alpha$-$MnO_2$, $\beta$-$MnO_2$, chemical manganese dioxide, CMD, and electrochemical manganese dioxide, EMD). The radii of the $MnO_2$ particles varied in the range 10...25 $\mu$m and their weight fraction 3%...30%. It was shown that the current density increased strongly with the decreasing particle size and scaled inversely with the particle radius, (1/$r$), which was attributed to an active triple phase boundary resulting from edge effects. In addition, OER overpotential depended greatly on the crystal structure of the $MnO_2$ in the composite. When compared with a similarly produced Pb-Ag anode, the electrode potential decreased as much as 250 mV in the short timescale (1 hour).

An expression for the total length of the boundary between catalyst particle and matrix metal was derived using cut statistics [17]. The length of the triple-phase boundary line (TPBL) was identified as the key variable explaining the experimental results. The TPBL has been also identified as a key property in solid oxide fuel cell studies [21–26], and for 2D electrodes the triple phase boundary (TPB) had a relation to the corrosion rate [27,28]. All these models assume an increased activity of the TPB leading to edge effects and to currents that are proportional to the length of the TPBL.

The mechanisms of the OER proposed in ref. [17] are illustrated in Figure 1. The first one (Figure 1a) is related to the ohmic drop within a particle during the OER (see inset of Figure 1a). The potential drop is lowest near the TPB, and the current density decays towards the centre of a $MnO_2$ particle. The second mechanism considers the transport of the OER products from the surface (Figure 1b). It assumes that the rate of oxygen evolution is reduced by increased surface concentration of protons or dissolved oxygen. Transport lowers their surface concentrations at the TPB and, therefore, increases the current density. The third mechanism is new, described as a "two-step-two-material mechanism" and involves hydrogen peroxide as an intermediate that is formed on the lead matrix and decomposed or oxidised on $MnO_2$ (Figure 1c). The last mechanism is characterised by oxygen bubbles covering most of the active $MnO_2$ particle surface, where only the TPB remains active (Figure 1d).

The mechanisms were studied with scanning electrochemical microscopy (SECM) and conductive atomic force microscopy (CAFM), targeting the activity of the interface between the metal matrix and the $MnO_2$ particle [29]. With CAFM, the special role of the TPB on the electrical properties was identified although electrochemical imaging of the active boundary was not achieved. SECM imaging showed that $MnO_2$ particles had a higher activity than the lead matrix. Scans in the Nernst diffusion layer for different species at varying tip potentials were interpreted via increasing concentrations of sulphuric acid and oxygen and decreasing concentrations of hydrogen peroxide towards the electrode. While increasing proton and oxygen concentrations were related to the process shown in Figure 1b, another hypothesis was that hydrogen peroxide is formed as an intermediate and reacts further on the composite electrode (Figure 1c); its formation on the counter electrode would be unlikely. However, due to missing topological information of the scanned area, it was not entirely clear if the SECM signal was due to topological contrast or electrochemical activity.

This communication has two goals: First, to generate random composite electrode surfaces and to verify the TPBL model; furthermore, to derive an expression for the active surface area of electrodes with protruding catalyst particles, in order to discriminate edge effects of the TPB from the surface area of the active particles. Second goal was to simulate three processes that may cause edge effects on the electrode. Thereby it should be possible to deduce which of the processes is the most probable.

The composite Pb-$MnO_2$ anodes studied in our earlier work [17] resemble randomly distributed arrays of ultramicroelectrodes (UMEs). The present study applies the ultramicroelectrode array model and diffusion domain approach of Amatore et al. [30] and Compton et al. [31–37] to model overlapping diffusional fields. In the case of randomly placed UMEs, the diffusion domain approximation involves Voronoi tessellation to calculate independent domain areas of the composite [37]. The domains are further made circular with a radius $r_0$, resulting in axisymmetric cylindrical domains that can be solved in 2D ($r,z$). For spatially heterogeneous or partially blocked electrodes simulations were able to accurately reproduce the experimental results [32,37].

Usually this approach is applied only to transient diffusion, but here it is used at steady-state to quantify edge effects resulting from the overlap of concentration and potential fields. Since both steady-state diffusion and electrical conduction are of the form of the Laplace equation, the approach is extended to solve the secondary current distribution, including the resistivity of the catalyst particles and the Pb matrix. It was assumed that the accuracy of the approximations made is sufficient to show trends in the current distribution as a function of the particle size, and allows to discriminate between the proposed mechanisms.

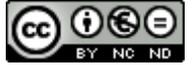

In addition to simulating different reaction mechanisms on Pb-MnO$_2$ composite anodes, the present work is of a more general nature. It studies the influence of geometric properties of composites on the efficiency of oxygen evolving anodes and might also be applied to coated titanium anodes where islands of IrO$_2$ are embedded in a Ta$_2$O$_5$ matrix [38].

## 2. Mathematical methods

All simulations were done with MATLAB 2015a (Mathworks, Natick, USA). The independent domain calculations were done in COMSOL 3.5a (COMSOL Inc, Burlington, USA) interfaced with MATLAB 2010a. A complete list of the used symbols within the article is found in Table 1.

## 2.1. Calculation of randomly spaced sphere packing

The mass fractions of the MnO$_2$ catalyst particles of interest were relatively high, over 5 wt%. With these mass fractions, randomly spaced points used to generate a distribution of the particles in the Pb matrix overlap. Therefore, using a Poisson type of distribution as in [37] does not model a random electrode surface properly, and the overlapping was gradually removed employing a technique that is known from molecular dynamics.

To simulate a pressed composite material, a Matlab code was developed employing a cube with periodic boundary conditions. The cube was filled with 10,000 rigid spheres of same size placed at random initial positions; the size of the cube depended on the mass fraction. The initial overlapping was removed during an iterative procedure. First, collision detection was done using a linked cell algorithm and, second, the forces between the particles were summed to result in the total force on each particle. The absolute value of the repulsive force $\|\vec{F}\|$ between two particles was calculated using Hertz's formula given in eqn. (2) of reference [39]. Attractive forces were not considered as they were expected to have an influence only on the nanoscale, whilst the particles in the composite are several orders of magnitude larger.

$$\|\vec{F}\| = (4/3)d^{3/2}\sqrt{R} \cdot E \tag{2}$$

where

$$R = \frac{r_1 r_2}{r_1 + r_2} \tag{3}$$

$$E = \frac{2E_1 E_2}{E_1 + E_2} \tag{4}$$

$E_1$, $E_2$ are the elastic moduli, $r_1$, $r_2$ the radii of each contact pair and $d$ is the depth of indentation or overlap.

For the time integration a velocity Verlet integrator [40] with an adaptive time step control was implemented. Each time step length was chosen so that the maximum displacement of a particle was 1% of its radius and the furthest moving particle determined the time step. The composite simulation was terminated when there were no remaining contacts between the particles.

## 2.2. Cutting of a simulated composite

To give a maximum statistical randomness each simulated composite cube was split randomly along all three orthogonal planes (XY, XZ, YZ), 3333 times for each plane, creating a cut-circle distribution in a square (see Figure 2). The cutting radius, $r_{cut}$, is the radius of the UME generated when cutting through the surface catalyst particles at certain height. The cut area through the cube was extended periodically by mapping the square to the opposite edges and corners. For this extended cut area a Voronoi tessellation was generated from the centre points of the cut circles. All areas with points that were not inside the "red cut square" of Figure 2b were rejected, and all Voronoi areas with vertices outside of the convex hull of the continued set of points were removed.

The remaining Voronoi areas were analysed by determining for each area the cutting radius, $r_{cut}$, and the independent "domain-area", $A_{domain}$ which was further made circular with a domain radius $r_0$. The resulting local coverage, $\theta$, is then given by equation (5).

$$\theta = \frac{A_{cut}}{A_{domain}} = \frac{\pi r_{cut}^2}{\pi r_0^2} = \left(\frac{r_{cut}}{r_0}\right)^2 \tag{5}$$

## 2.3. Diffusion domain approach

The method of the diffusion domain approach is similar to the modelling of partially blocked electrodes by Compton et al. [37] but uses the methods described above to derive the distribution function related to the UME size, $r_{cut}$ and the local coverage, $\theta$. After their calculation, a histogram of the domain area sizes is created spanning the intervals $r_{cut} = r_n = [0, r_p]$ and $\theta = [0,1]$, resulting in the count matrix $N(r_{1...n}, \theta_{1...m})$. Thereafter, the corresponding currents $i(r_n, \theta_m)$ are simulated in cylindrical coordinates. Finally, the total current through the composite can be calculated according to formula:

$$I = \sum_n \sum_m N(r_n, \theta_m) \cdot i(r_n, \theta_m) \tag{6}$$

The calculation of the total current of a composite requires ca. 2000 simulations.

The procedure is illustrated in Figure 3, where it is carried out for different domain sizes with constant cutting radii. The first step is to generate a random composite electrode surface, which is achieved by using the methods outlined above. Second, the independent domains are found using the Voronoi tessellation [33] which generates lines at the half-distance between the points. A random electrode surface is illustrated by a contour plot of surface concentrations in Figure 3a, where a Voronoi tessellation (white lines) was used to split the surface into independent domains. Next, the surface area of Voronoi polygons was determined. The share of the domain sizes in Figure 3a of the dimensionless area 8.5-9.5(9), 9.5-10.5(10), 10.5-11.5(11) and 11.5-12.5(12) are sketched in Figure 3b. After that cylindrical domains of the base area size 9, 10, 11 and 12 (seen in Figure 3c) are simulated, multiplied by their share and summed up to give the total current of the array.

As it is common for unknown convection profiles, the equation of convective diffusion is approximated with the plain diffusion equation, assuming a Nernst diffusion layer of the thickness $\delta$. This approximation is also used in the diffusion domain approach simulations and applied here with a diffusion layer thickness $\delta = 100\ \mu m$. In contrast to the Nernst diffusion layer approximation, which is based on efficient stirring and leads to uniform concentrations in the bulk electrolyte, the potential drop in the bulk electrolyte is not zero. Due to the nature of the diffusion domain approach it had to be assumed that the potential at the distance $z = 100\ \mu m$ from the electrode is equal across the whole electrode. This is often the case in common electrode setups, since even with local heterogeneity; the current density and the potential are homogeneous at large distances from the electrode.

## 2.4. Models

This section describes the boundary value problems that are solved to calculate the total current of a composite electrode according to equation (6). The model of mechanism 1 (Figure 1a) is described in section 2.4.1, that of mechanism 2 (Figure 1b) in section 2.4.2 and that of mechanism 3 (Figure 1c) in section 2.4.3. All three mechanisms address different aspects that can lead to edge effects, *viz* to an active triple phase boundary. Although some of the processes can occur simultaneously, each model, combined with the diffusion domain approach, serves the purpose to evaluate only the effects related to the specific mechanism. This means that the model of mechanism 1 solves only the secondary current distribution, and effects arising from concentration gradients are not taken into account. Mechanism 2 targets only the mass transfer of protons and its influence on the rate of the OER. Therefore, the potential drop in the electrolyte and electrode is not included in the calculation of the potential dependence of the rate constant. Similarly, the model of mechanism 3 solves only the diffusion equation of hydrogen peroxide subjected to kinetic boundary conditions, where the rate constants depended directly on the applied potential.

### 2.4.1. Secondary current distribution

The OER is assumed to proceed via the electrochemical oxide path with the 1$^{st}$ step rate determining; the reaction mechanism and the derivation of its rate are given in the supplementary material. To investigate the effect of the low conductivity of $MnO_2$ on the current distribution, the following boundary value problem was solved (see also Figure 4). Superscripts 1, 2 and 3 denote hereafter the electrolyte ($H_2SO_4$), $MnO_2$ and Pb, respectively. The electric potential is solved from electrostatics in the three domains:

$$\nabla \cdot (\sigma \nabla \phi_k) = 0 \rightarrow \nabla^2 \phi_k = 0 \; ; \; k = 1, 2, 3 \tag{7}$$

The potential $\phi_1$ in the electrolyte solution should be calculated with the Nernst-Planck equation but in the secondary current distribution concentration polarization is omitted, justifying the use of the Laplace equation.

The potential is coupled with the boundary condition

$$\sigma_2 \left(\frac{\partial \phi_2}{\partial z}\right)_{z=0} = \sigma_1 \left(\frac{\partial \phi_1}{\partial z}\right)_{z=0} = 4Fk^0 \, e^{(\alpha F/RT)(\phi_2 - \phi_1 - E^0)}; \quad 0 < r < r_{cut} \tag{8}$$

Equation (8) is the kinetic equation of oxygen evolution, reaction (1), neglecting the inverse reaction. It also represents the current continuity at the solution/$MnO_2$ boundary. The current continuity between Pb and $MnO_2$ is given by equation (9), where $\vec{n}_i$ denotes the surface normal of the respective boundary:

$$\vec{n}_2 \cdot \vec{j}_2 - \vec{n}_3 \cdot \vec{j}_3 = 0 \tag{9}$$

The boundary between lead and the electrolyte is assumed to be insulating

$$\left(\frac{\partial \phi_k}{\partial z}\right)_{z=0} = 0 \; ; \; k = 1, 3; \; \; r_0 < r < r_{cut} \tag{10}$$

The symmetry condition at the centre axis and the insulation condition at the domain boundary reads

$$\left(\frac{\partial \phi_k}{\partial r}\right)_{r=0} = \left(\frac{\partial \phi_k}{\partial r}\right)_{r=r_0} = 0 \; ; \; k = 1, 2, 3; \; \; 0 < z < \delta \tag{11}$$

Finally, the electric potential in the solution "far" from the electrode is zero:

$$\phi_1\,(z = \delta, 0 < r < r_0) = 0 \tag{12}$$

And the potential applied to the bottom of the simulation domain is $E$:

$$\phi_3\,(z = -r_0, 0 < r < r_0) = E \tag{13}$$

The conductivity of sulphuric acid (2M, 25°C), $\sigma_1$ was taken as 70 S/m [41] and the conductivity of the MnO$_2$ particles, $\sigma_2$, as 0.1, 1 or 10 S/m; for Pb the literature conductivity value, $\sigma_3$ = 4.8×10$^6$ S/m, was used. An apparent standard rate constant, $k^0$ = 1.3×10$^{-7}$ mol·m$^{-2}$s$^{-1}$ and the charge transfer coefficient, $\alpha$ = 0.5, were found to agree with the previous experimental results. To include the effect of the ohmic drop, the potential difference, $\phi_2 - \phi_1$ between the electrolyte and the electrode surface was applied as the driving force, representing the applied potential at a given position on MnO$_2$. Particle radii and the electrode potential were varied at a constant global surface coverage of the MnO$_2$ particles, $\Theta$ = 0.25.

### 2.4.2. Mass transfer of protons

This model describes the accumulation and transport of sulphuric acid, as sketched in Figure 1b. In our previous paper [17] pure sulphuric acid was used, and only a small amount of the bisulphate dissociates into sulphate. Hence, the sulphate species is neglected, which greatly simplifies the calculation. The advantage of having only two ions (proton and bisulphate) is that, due to electroneutrality, their concentrations are equal to that of sulphuric acid, $c_{H^+} = c_{HSO_4^-} = c_{H_2SO_4} = c_1$. The Nernst-Planck equations can now be simplified to the diffusion equation of a binary electrolyte [42] where the migration terms are eliminated, as shown in the supplementary material.

The model also assumes that the rate of oxygen evolution is reduced by an increased proton concentration, which is reflected in its apparent reaction order of −1:

$$j = 4Fk^0 c_{H^+}^{-1}\, e^{(1+\alpha_2)F/RT(E-E^o)} \tag{14}$$

Equation (14) is derived in the supplementary material, following the treatise by Bockris [43] and assuming the electrochemical oxide path with the second step rate determining.

The simulated domain with modified boundary conditions, equations (15)-(19), is shown in Figure 5. Diffusion of sulphuric acid diffusion at steady-state:

$$\nabla \cdot (D_1\,\nabla c_1) \to \nabla^2 c_1 = 0 \tag{15}$$

The reaction rate on the MnO$_2$ particle is given by equation (16)[1]:

$$\left(\frac{\partial c_1}{\partial z}\right)_{z=0} = -\frac{4k^0}{2D_{H^+}} c_1^{-1} \cdot e^{(1+\alpha_2)\frac{F}{RT}(E-E^o)}; \quad 0 < r < r_{cut} \tag{16}$$

Again, it is assumed that no reaction occurs on the Pb matrix:

$$\left(\frac{\partial c_1}{\partial z}\right)_{z=0} = 0; \quad r_0 < r < r_{cut} \tag{17}$$

On the symmetry axis and on the domain boundary the gradient is zero:

---

[1] With the definitions of the proton transport number $t_+$ and $D_{H_2SO_4}$, and employing the Nernst-Hartley relation, it follows that in the boundary condition $\frac{D_{H_2SO_4}}{1-t_+} = 2D_{H^+}$.

$$\left(\frac{\partial c_1}{\partial r}\right)_{r=0} = \left(\frac{\partial c_1}{\partial r}\right)_{r=r_0} = 0; \quad 0 < z < \delta \tag{18}$$

At the limit of the simulation domain the concentration of sulphuric acid is set to its bulk value:

$$c_1(z = \delta, 0 < r < r_0) = c_1^{bulk} \tag{19}$$

In order to account for non-ideality, a measured average diffusion coefficient of sulphuric acid of $D_1$ = $2.2 \times 10^{-5}$ cm$^2$/s [41,44] and the literature value for the diffusion coefficient of protons, $D_{H^+}$ = $9.3 \times 10^{-5}$ cm$^2$/s [42], were used. The bulk concentration of sulphuric acid, $c_1^{bulk}$, was set to 2 M and the diffusion layer thickness to 100 μm. The charge transfer coefficient of step 2, $\alpha_2$ was 0.5 and the matching apparent standard rate constant, $k^0$, was determined to be $9 \times 10^{-15}$ mol$^2$m$^{-5}$s$^{-1}$.

### 2.4.3. Two-step-two-material mechanism

This model describes the reactions of the hydrogen peroxide and diffusion in the cylindrical domain. It is assumed that PbSO$_4$ or PbO$_2$ residing on the surface generates H$_2$O$_2$ via reaction (20), followed by its decomposition or oxidation on the MnO$_2$ catalyst, reaction (21) (see also Figure 1c).

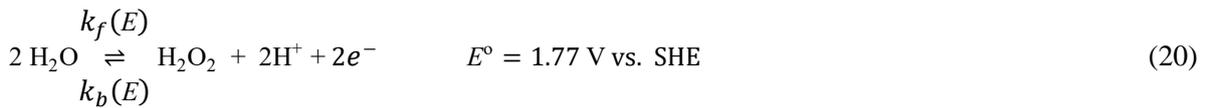

$$2\ H_2O \underset{k_b(E)}{\overset{k_f(E)}{\rightleftharpoons}} H_2O_2 + 2H^+ + 2e^- \qquad E^\circ = 1.77\ V\ vs.\ SHE \tag{20}$$

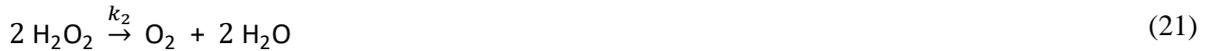

$$2\ H_2O_2 \overset{k_2}{\rightarrow} O_2 + 2\ H_2O \tag{21}$$

In order to study the influence of the decomposition reaction (20) on the generation of hydrogen peroxide, reaction (20) was assumed to be quasi-reversible. It was also assumed that no decomposition of H$_2$O$_2$ occurred on the lead surface. A fraction of H$_2$O$_2$ would decompose in the solution, but that is also neglected because H$_2$O$_2$ is relatively stable in sulphuric acid solutions.

As the purpose of the simulation was to see if the decomposition of H$_2$O$_2$ on the MnO$_2$ catalyst could enhance the formation of H$_2$O$_2$ on the Pb matrix, there was no attempt to derive an exact mechanism involving surface intermediates. Instead, a simplified reaction mechanism is assumed: The first step of reaction (20) is treated as a two-electron electron transfer reaction following Butler-Volmer kinetics. The forward and backward rate constants, $k_f$ and $k_b$ respectively, are given by equations (22) and (23):

$$k_f(E) = k^0 \exp\left(\frac{2\alpha F}{RT}(E - E^\circ)\right) \tag{22}$$

$$k_b(E) = k^0 \exp\left(-\frac{2(1-\alpha)F}{RT}(E - E^\circ)\right) \tag{23}$$

The second step, hydrogen peroxide decomposition, is treated as a heterogeneous, first order irreversible reaction with the rate constant $k_2$. For the backward reaction of (20), a reaction order of 1 was assumed with respect to H$_2$O$_2$. The activity of water was set to unity. The influence of the proton concentration was neglected, as pH was constant (~0).

The resulting boundary value problem is represented in Figure 6 and is given by equations (24)-(28). The bulk solution is described by the diffusion equation of hydrogen peroxide:

$$\nabla \cdot \left(D_{H_2O_2} \nabla c_{H_2O_2}\right) \rightarrow \nabla^2 c_{H_2O_2} = 0 \tag{24}$$

Reaction (20) taking place on the lead matrix is described by boundary condition (25):

$$D_{H_2O_2}\left(\frac{\partial c_{H_2O_2}}{\partial z}\right)_{z=0} = -k_f + k_b c_{H_2O_2}; \quad r_{cut} < r < r_0 \tag{25}$$

The decomposition of $H_2O_2$ on the $MnO_2$ catalyst particle results in boundary condition (26):

$$D_{H_2O_2}\left(\frac{\partial c_{H_2O_2}}{\partial z}\right)_{z=0} = k_2 c_{H_2O_2}; \quad 0 < r < r_{cut} \tag{26}$$

Symmetry and domain boundary conditions are

$$\left(\frac{\partial c_{H_2O_2}}{\partial r}\right)_{r=0} = \left(\frac{\partial c_{H_2O_2}}{\partial r}\right)_{r=r_0} = 0; \; 0 < z < \delta \tag{27}$$

Additionally, an insulating boundary condition in the bulk solution was chosen because at steady-state, all $H_2O_2$ generated would be decomposed on the $MnO_2$ catalyst. Initial simulations showed that at steady-state the concentration of $H_2O_2$ levelled out at further distance from the electrode. Therefore, the height of the simulation domain was limited to twice the domain radius.

$$\left(\frac{\partial c_{H_2O_2}}{\partial z}\right)_{z=\delta=2r_0} = 0; \quad 0 < r < r_0 \tag{28}$$

For the simulation the following parameter values were chosen: $D_{H_2O_2}$ = 1.8×$10^{-5}$cm$^2$/s [29,45], $\delta$ = 2$r_0$ = $2r_{cut}/\sqrt{\theta}$, and, if not stated elsewhere, a volume fraction of $\Theta$ = 0.25 and a particle radius of $r_p$ = 10 μm. The electrochemical rate constant $k^0$ of $10^{-3}$cm s$^{-1}$ was found to fit with the experimental results reported in ref. [17].

# 3. Results

## 3.1. Validation of triple phase boundary length model

The model presented in [17] describes the total length of the boundaries between the surface catalyst particles ($MnO_2$) and the surface metal matrix (lead), i.e. the triple phase boundary between the catalyst particles, the metal matrix and the electrolyte. When this boundary is more electrochemically active than the surrounding composite, the current density increases with the length of the boundaries per unit area. Thus, not only the true surface area, but also the TPBL is a potentially important geometrical property of composite electrodes.

The number of particles, $N$, residing on the surface and the total length, $L$, of the triple phase boundary are given in equations (29) and (30); their derivation is given in the supplementary material.

$$N = \frac{\Theta \cdot A_{geom}}{\bar{A}_{cut}} = \frac{3}{2\pi} \cdot \frac{\Theta \cdot A_{geom}}{r_p^2} \tag{29}$$

$$L = \sum l_i = 2\pi \sum r_{cut,i} = N \cdot \bar{l} = 2\pi \cdot N \cdot \bar{r}_{cut} = \frac{3}{4}\pi \cdot \frac{\Theta \cdot A_{geom}}{r_p} = \frac{3}{4}\pi \cdot \frac{vol\% \cdot A_{geom}}{r_p} \tag{30}$$

Figure 7a shows that the simulated TPBL values are proportional to $1/r$ and also to the volume fraction. Furthermore, Figure 7b shows that the TPBL values calculated from equation (30) and the calculated values of the TPBL obtained from the cut simulation were identical; hence, the mathematical model for the TPBL and its simulation agree.

### 3.1.1. Active surface area of compact and porous electrodes

The active surface area of common flat electrodes can be determined in a similar manner and it appears to be independent of the particle radius, see equation (31).

$$A = \sum A_i = N \cdot \bar{A} = \Theta \cdot A_{geom} = vol\% \cdot A_{geom} \qquad (31)$$

It is known that the surface fraction or global coverage, $\Theta$ and volume fraction, *vol%* are the same [46], which is also proven in the supplementary material. Even if the catalyst particles were protruding from the supporting matrix, the total active surface area of the composite electrode would be independent of the particle radius. This can be shown by calculating the average surface area of a sphere cap:

$$\bar{A}_{exposed} = \frac{1}{2r_p} \int_0^{2r_p} 2\pi r_p h \, dh = 2\pi r_p^2 = \frac{1}{N} \sum A_{exposed,i} \qquad (32)$$

Since the number of active particles at the surface remains the same as for the flat electrode, the total active surface area of a composite with protruding spherical particles is obtained by equation (33):

$$A = \sum A_{exposed,i} = N \cdot \bar{A}_{exposed} = 2\pi r_p^2 \frac{3}{2\pi} \cdot \frac{\Theta \cdot A_{geom}}{r_p^2} = 3\Theta A_{geom} \qquad (33)$$

Thus, even for a composite electrode with protruding particles, the active surface area is independent of the radius; and the factor 3 can be considered to be a roughness factor which will be different for other non-flat surfaces.

In porous electrodes, nearly all the particles are active and the number of active particles scales with $1/r_p^3$. Multiplying it with an average active surface area at the internal surface that is proportional to $r_p^2$, the total active surface area of a porous electrode scales with $1/r_p$. However, since composite electrodes that have a metal matrix, are classified as compact [5], only the particles residing on the surface are active. Therefore an active triple phase boundary and edge effects are more likely the cause of the observed experimental results in ref. [17] than an increased surface area. The TPBL formula (30) could thus explain the geometric origin of the experimentally observed results, but it does not explain why the triple phase boundary is more active than the rest of the composite. The results of the models introduced in sections 2.4.1-2.4.3 allow a more detailed evaluation.

## 3.2. Evaluation of proposed processes

### 3.2.1. Influence of particle resistivity on secondary current distribution

Figure 8a shows the effect of the conductivity of the $MnO_2$ particles on the current density as a function of the particle radius ($r_p$). The electrode potential was $E - E° = 0.6$ V. As it can be seen, the smaller $r_p$ is, the smaller is the influence of the ohmic drop. Figure 8b shows a simulation of the effect of the electrode potential. The conductivity of the $MnO_2$ catalyst particles, $\sigma_2$, was kept constant at 1 S/m. The current density is scaled with $1/r_p$ and the influence of $r_p$ increased with increasing electrode potentials, similar to the experimental results in Figure 8d. Figure 8c shows that local current density increases towards the TPB at varying electrode potentials. The geometric current density is approximately 25% of the average local current density at a $MnO_2$ catalyst particle when the catalyst covers only 25% of the surface.

On the basis of Figure 8a-c, it can be proposed that current from the lead matrix to a catalyst particle into the electrolyte flows near the TPB because this represents the shortest path (see inset of Figure 1a), explaining why the measured current density in Figure 8d scales with $1/r_p$. Another similarity between the model (Figure 8b) and the experimental results in Figure 8d is that an increased electrode potential also leads to a steeper increase of the current density with $1/r_p$. However, with the model, edge effects were less pronounced with decreasing $MnO_2$ catalyst particle size which leads to bent slopes of $J$ vs. $1/r_p$. Such behaviour was not observed experimentally.

### 3.2.2. Influence of mass transfer of acid

The simulation of current density due to the transport of protons at an electrode with 25 vol% of catalyst particles is presented in Figure 9a as a function of particle radius. The plot shows that the catalyst particle size had no effect on the current density even when the apparent reaction order is −1. Figure 9b also shows that the current density on a 10 μm particle in a 20μm domain was uniform and that no edge effects occurred.

### 3.2.3. Two-step-two-material reaction

The effect of the rate constant $k_2$ on the current density, Tafel slope and on the steady-state bulk concentration of $H_2O_2$ was studied by varying the electrode potential, taking a relatively high electrochemical rate constant, $k^0 = 10^{-3}$ cm s$^{-1}$. Simulations in Figure 10a show that increasing $k_2$ increased also the current density. The Tafel slope in Figure 10b changed from 30 mV to ca. 60 mV, but decreased again at higher potentials. The bulk concentrations shown in Figure 10c followed the Nernst equation for low values of $k_2$ and/or low potentials, but were lower at higher potentials with high values of $k_2$.

This can be related to a competition between the backward reaction of (20) and the reaction (21). In the case of $k_b(E) \gg k_2$ the equilibrium concentration calculated from the Nernst equation was obtained. Similarly to the equilibrium concentrations, the bulk concentrations increased by an order of magnitude with an increase of the electrode potential by each 30 mV, which is the same value as the Tafel slope. In the opposite case, where the decomposition of $H_2O_2$ was much faster than the reverse reaction of (20) – at potentials where $k_2 \gg k_b(E)$ – the bulk concentration is significantly lower than the equilibrium concentrations calculated from the Nernst equation, which could be a sign of that most of the generated $H_2O_2$ decomposed at the catalyst. At the same time the Tafel slope changed to 60 mV, which is another indication of the transition to an irreversible process where the forward reaction rate of (20) is rate determining.

The influence of the $MnO_2$ particle size on the current density was studied at different applied potentials with the rate constant $k_2 = 10$ cm s$^{-1}$ and a 25% volume fraction of particles. Experimental results from ref. [17] are shown in Figure 11b for comparison. As can be seen, the results of the simulation correlate quite well with the experimental results. The simulated current density scaled with $1/r_p$ and the effect increased with increasing potential. Yet, the model showed a weaker influence of the particle radius with smaller radii, which is not seen from the experimental results.

To visualise the influence of the reverse reaction (20) on the current distribution at the TPB, the radial distribution of the current density and the surface concentrations at the lead component were plotted as a function of the particle radius, normalised by the domain size. The electrode potential, $E - E^0$, was −20 mV. The radial current density (Figure 11c) increased towards the TPB which can be explained by the lower concentrations of $H_2O_2$ seen in Figure 11d. Since the reverse reaction on the Pb matrix is hindered, the current increased at the TPB.

Figure 11 shows that small catalyst particles were the most active in the decomposition of $H_2O_2$. Consequently, with smaller particles less catalyst is needed. Since both the lead matrix and $MnO_2$ catalyst particles are involved in the reaction, the mechanism was studied for different volume fractions of catalyst. In the simulations of Figure 12a, with a particle radius of 10 μm, it was observed that the current density was going through a maximum when increasing the $MnO_2$ surface coverage. Figure 12b shows a corresponding plot of $J_{max}$ vs. the surface coverage. At low potentials (current densities) the optimum volume fraction of the catalyst was greater than at high potentials (current densities). Increasing the volume fraction of $MnO_2$ above the optimum amount did not have a large effect, but for higher potentials (current densities) an excess of catalyst caused a significant decrease in the current density, which could relate to a reduction of generation area. It also shows that the amount of $MnO_2$ must be chosen carefully to obtain optimal results.

If the decomposition reaction on $MnO_2$ would be replaced by an irreversible first order electrooxidation reaction instead, the reaction rate constant $k_2$ would change to a potential dependent rate constant $k_{ox}(E)$. As a consequence, the current-potential behaviour would change. Yet the current densities obtained during the simulation with different rate constants $k_2$ (in Figure 10a) can be matched with a corresponding $k_{ox}(E)$ where the current density would need to be multiplied by two to account for the current flowing through the $MnO_2$ catalyst particle. Edge effects in current density found on the $H_2O_2$ generating area (Figure 11c) would extend to the catalyst particle area.

## 4. Discussion

The first aspect to discuss is the role of the TPBL, or in particular, if the experimental results are due to edge effects caused by an active TPB, or due to a porous layer where almost all catalyst particles are active (cf. section 3.1.1). In our earlier work [17] bulk metal matrix composites were studied. These contained deeply buried catalyst particles, and since the whole pressed tablet contained $MnO_2$ it is not plausible that all the particles would be active. Because the existence of an active TPB was a hypothesis on which all the simulations were based, it was important to show that in non-porous composites the active surface area does not increase with smaller particles or at least the active surface area does not scale with $1/r_p$.

The simulations of the mass transport of sulphuric acid showed that only very gradual concentration gradients of acid were found. This can be explained by the migration of protons carrying away most of the generated acid. Therefore, no edge effects were present and the current density did not depend on the particle size. This finding contrasts with the SECM results [29] where it was concluded that the sulphuric acid concentration increased substantially towards the composite electrode surface. The origin of these results is thus not yet fully understood and should be reviewed with a similar methodology as in section 2.4.2, but involving convection. Unless extreme current densities really do occur, mass transfer of acid is not likely to be the explanation of the experimentally obtained $1/r_p$ relationship with the current density in ref. [17].

The other two simulated mechanisms, however, showed edge effects that are large enough to explain the experimental results and the dependency of the current density on the particle size. Yet, there is a notable difference in the plots of the (geometric) current density versus $1/r_p$ between measurement and simulation, implying more pronounced edge effects in the real mechanism. Nonetheless it remains unclear if this indicates a completely different process from those simulated in this work, or if it could be, e.g. a result of a two-step two-material mechanism with electrooxidation of $H_2O_2$, which combines the effects of a potential drop in $MnO_2$ with the characteristics of mechanism 3.

Anyway, the simulations showed that the catalytic activity of $MnO_2$ is hampered when current flows through thick coatings or large particles. Also, it could be concluded that for resistive particles incorporated into a well conducting matrix, it would depend on the particle size, whether pronounced edge effects would occur or if the catalyst particles are small enough to keep the iR drops small.

The two-step-two-material mechanism, is, however, rather complex, even when assuming a simplified reaction path. Regarding the simulations, a reader may question especially the very high value of the decomposition rate constant, $k_2 = 10$ cm/s. It does not depend on the electrode potential, but if instead of decomposition, electrooxidation of $H_2O_2$ is considered, the rate constant would relate to reaction (35) which has a very low standard electrode potential. Therefore, reaction (35) would occur at large overpotentials. More problematic is the formation of $H_2O_2$: the apparent standard rate constant $k^0 = 10^{-3}$ cm/s of reaction (34) is relatively high for a multistep inner-sphere electron transfer.

Since the presence of $H_2O_2$ in oxygen reduction is evident, see e.g. references [45,47], it can be considered as a plausible intermediate in the OER. Such a mechanism has been proposed by Newman and Alyea [48] whereby $H_2O_2$ forming via reaction (34) would more or less instantly oxidize to oxygen via reaction (35). This would mean that $H_2O_2$ should be present only in very small amounts.

$$2\,H_2O \rightleftharpoons H_2O_2 + 2H^+ + 2e^- \qquad E^o = 1.77\text{ V vs. SHE} \qquad (34)$$

$$H_2O_2 \rightarrow O_2 + 2H^+ + 2e^- \qquad E^o = 0.65\text{ V vs SHE} \qquad (35)$$

The formation of hydrogen peroxide during the OER has been discussed by Fierro et al [49]. They classified OER electrocatalysts into two categories: the first one involving weakly physisorbing hydroxyl radicals and hydrogen peroxide as intermediates, and the second one characterised by chemisorbed hydroxyl radicals and higher oxides. Only the latter type was identified as a good electrocatalyst. The presence of $H_2O_2$ was experimentally investigated by Pavlov and Monahov who claimed that $H_2O_2$ is not involved as an intermediate during oxygen evolution on lead anodes [50]. Nonetheless, they suggested that adsorbed hydroxyl radicals are formed as intermediates. Other experimental studies, apart from ref. [17], that target soluble intermediates during the OER are scarce. Kuznetsova et al. [51] showed that even for electrocatalysts with strong adsorption of water ($IrO_2$ and $RuO_2$), unidentified soluble intermediates formed during oxygen evolution, but they were observed only under dynamic conditions.

On the basis of Gibbs free energies of adsorption, generation of hydrogen peroxide on metal oxides has been recently suggested by Viswanathan et al. [52]. They claimed that with materials having weak adsorption of water, that usually do not show good catalytic properties in the OER, hydroxyl radicals should desorb into the solution to form hydrogen peroxide, instead of being oxidised further to oxygen. The Gibbs free energy of formation of the adsorbed intermediates on crystalline β-$PbO_2$ has been calculated by Mom et al. [53] and it corresponds to a range where the formation of hydrogen peroxide should be favoured [52]. The formation of the hydroxyl radicals on β-$PbO_2$ is, however, still characterised by a relatively high energy barrier which could mean that additional overpotential would be needed to form $H_2O_2$ in substantial amounts. Therefore, further experimental evidence for $H_2O_2$ formation is needed.

Although $H_2O_2$ was detected in an earlier work using SECM [29], conclusions regarding the existence of a two-step-two-material mechanism on that basis are not easily drawn. It was not possible to image the formation of $H_2O_2$ on the composite electrode, neither was the triple phase boundary accessible. Alternatively, it should be possible to study the electrochemical reactions of $H_2O_2$ formation, decomposition and electrooxidation on Pb and $MnO_2$ electrodes and to deduce whether measured rate constants would support the existence of this reaction mechanism.

The third possible mechanism, the coverage of the electrode by gas bubbles, could be seen as another reason of edge effects (Figure 1d). It seems reasonable that only when bubbles reside on the active $MnO_2$ catalyst particles, though not completely covering all active area, a correlation between the catalyst particle size and the current density can be obtained. For homogeneous electrode materials Dukovic and Tobias simulated the current distribution around a gas bubble attached to an electrode surface, taking ohmic drops, activation overpotential and concentration overpotential into account [54]. They concluded that for hydrogen evolution in chloralkali electrolysis cells, the current density is distributed evenly on areas that are not covered by bubbles. It is thus to be expected that bubbles residing on the composite electrode would predominantly reduce the available surface area. Bubbles residing on inactive parts of the composite will thus barely influence the current distribution, while bubbles on active areas will simply block the reactive area and reduce the overall current density. How this would exactly affect each of the presented processes, is however, not exactly clear.

If the bubble coverage extends randomly over both materials, gas evolution on the composite can be treated as a process taking place on a single effective material, as in the Cassie-Baxter wetting model [55,56]. Knowledge of the actual bubble coverage would clarify the situation and could also serve as another tool to further optimise the composite for oxygen evolution.

There is yet another mechanism presumably leading to edge effects. Adsorbed surface intermediates may block active sites. However, when these intermediates diffuse across the triple phase boundary to the catalyst - a phenomenon known as a "spill-over" mechanism [57] - and react further, a constant flow of intermediates towards the catalyst is observed, re-exposing active sites on the generation area and increasing the total rate in a similar manner to the two-step-two-material mechanism.

## 5. Conclusion

Electrochemical reactions on composite electrodes were simulated using an extended diffusion domain approach employing a method similar to Davies et *al* [32]. Composite electrode surfaces with randomly distributed catalyst particles were simulated using a force based sphere packing algorithm, followed by cutting to expose a composite electrode surface. The simulated composite material surfaces were used to generate diffusionally independent domains based on the Voronoi tessellation. 2D histograms were calculated using the local coverage and the cut radii of the domains. Each domain shape in the histogram was simulated in cylindrical coordinates and, finally, the total current was calculated as a weighted sum.

The mathematical model predicting the triple phase boundary length (TPBL), that was proposed to explain the previous experimental results in [17], was verified. The usual assumption of the relationship between the particle radius and the surface area is that composite electrodes containing small particles have a larger total surface area than those with larger particles. For compact 2D composite electrodes [5] used in our previous studies [17], this assumption does not necessarily hold true. Especially, if most of the particles are buried in the composite - and the surface coverage percentage is the same as the volume percentage - the total active surface area should be independent of the particle radius.

It was shown that two of the three simulated electrochemical microscopic process models showed promising results matching the behaviour of an active triple phase boundary: i) the influence of the catalyst conductivity on the secondary current distribution and, ii) the two-step-two-material mechanism with quasi-reversible generation of the intermediate $H_2O_2$ on the lead matrix followed by its decomposition on $MnO_2$. The mass transfer of protons did not seem to fit to the experimental results and did not produce any increased activity at the triple phase boundary. Screening by gas bubbles on the electrode has obviously an influence on the system. This was not simulated here, but it was suggested that only bubbles on top of the $MnO_2$ particles would cause edge effects in the local current density.

The influence of the conductivity explains why a catalytically active but poorly conducting catalyst performs well when combined with a conducting matrix that binds well to the catalyst. This enables the use of electrocatalysts in composite materials which would otherwise be considered as having too poor a conductivity.

Proving the existence of the two-step-two-material mechanism is more challenging. Theoretical considerations of Viswanathan et al. [52] suggest that the formation of hydrogen peroxide on lead is kinetically favoured although, considering the adsorption energies on $\beta$-$PbO_2$ calculated by DFT [53], it might be relatively slow. In the present simulations, high values of the rate constants were required to fit the experimental results and, therefore, further experimental evidence of the existence of this mechanism is needed. Otherwise this work shows that, giving the right materials, the two-step-two-material mechanism can be used to tailor metal matrix metal oxide (MMMO) composite anodes or mixed metal oxide (MMO) composite anodes and to tune the components separately. The first catalyst component would be selected to form $H_2O_2$ (or another intermediate) and the second one would be optimised for its decomposition or electro-oxidation. This could be a means to find cost effective oxygen evolving anodes with relatively low overvoltages.

## *Acknowledgements*

Kai Vuorilehto is acknowledged for the input to total surface area calculations with 3D electrodes. Late Professor Kyösti Kontturi (1949 - 2015) is remembered for his advice and support to this work.

## *Tables*

Table 1        List of Symbols

| Notation | Unit/Value | Description |
|---|---|---|
| $E^\ominus$ | V | Standard electrode potential |
| $\|\vec{F}\|$ | N | Absolute value of the force in between two touching spheres |
| $d$ | µm | Overlap in between two touching spheres |
| $E, E_1, E_2$ | MPa | (Effective) elastic modulus |
| $R$ | µm | Effective radius |
| $r_p, r_1, r_2$ | µm | Particle radii |
| $I, i, i(r_i, \theta_j),$ | A, nA | Total current, current in one domain, current for a specific domain size and local coverage |
| $\theta, \theta_j, \theta_{OH}, \theta_O$ | 0…1 | Local coverage, surface fraction of $MnO_2$ in one Voronoi domain, coverage by the OER intermediate MOH respective MO |
| $N(r_i, \theta_j), N$ | | Number of particles with specific domain size and local coverage, number of cut particles |
| $\phi$ | V | Potential |
| $J$ | A/m$^2$ | Current density |
| $\vec{n}$ | | Surface normal (on a boundary) |
| $k^0$ | | (apparent) standard reaction rate constant |
| $\alpha$ | 0…1 | Charge transfer coefficient |
| $F$ | 96485 As/mol | Faraday constant |
| $R$ | 8.3145 J/molK | Ideal gas constant |
| $T$ | K | Temperature |
| $E$ | V | Electrode potential |
| $z, r, r_{rad}$ | µm | Height, radial coordinate (cylindrical coordinates) |
| $\delta, r_p, r_0$ | µm | Nernst diffusion layer thickness, particle radius, domain radius |
| $c, c_{H^+}, c_{HSO_4^-}, c_{H_2SO_4}, c_{H_2O_2},$ $c_1^{bulk}$ | mol/L | Concentrations of the species, bulk concentration of sulphuric acid |
| $D_{H^+}, D_{HSO_4^-},$ $D_{H_2SO_4}, D_{H_2O_2}$ | cm$^2$/s | Diffusion coefficients of the species |
| $k_f, k_b, k_2$ | cm/s | Forward and reverse reaction rate constant, $H_2O_2$ decomposition rate constant |
| $r_n, r_{cut}, \bar{r}_{cut}, h$ | µm | Radius of cut particle $n$, average cutting radius, cutting height |
| $\Theta$ | 0…1 | Global/average surface coverage by catalyst ($MnO_2$) |
| $vol\%$ | 0…1 | Composite volume fraction of catalyst ($MnO_2$) |
| $A, A_{geom}, A_i, \bar{A}_{cut},$ $A_{exposed,i}, \bar{A}_{exposed}$ | m$^2$, µm$^2$ | Total surface area, geometric electrode surface area, cut surface area($MnO_2$) in domain $i$, average cut surface area, exposed surface area in domain $i$, average exposed surface area |
| $L, l_i, \bar{l}$ | µm, m/m$^2$ | Length of the triple phase boundary (TPBL), length for domain $i$, average length for one domain |
| $J_{geom}, J_{max}$ | A/m$^2$ | Current density with respect to geometric surface area, Maximum in geometric current density (with respect to the volume fraction) |

## *Figure captions*

Figure 1. Four mechanisms proposed in ref. [17] leading to an active TPB. a) Electron transfer through the surface particles. b) Transport of products having an adverse effect on the reaction rate. c) Mechanism involving $H_2O_2$ as an intermediate that forms on Pb and reacts further on $MnO_2$. d) Growth of gas bubbles on $MnO_2$, screening all active material apart from the TPB [17].

Figure 2. a) Cutting through the distribution in a cube containing 1000 spheres (25 vol% of spheres). b) Generation of a Voronoi tessellation on an extended cut area.

Figure 3. Illustration of the diffusion domain approach at a UME array surface at the limiting current. a) Surface concentrations for randomly spaced microelectrodes, independent domains determined by Voronoi tessellation. b) Histogram of the domain area sizes. c) Diffusion domains used to simulate total current of the UME array. Numbers indicate weighting factors for summation.

Figure 4. Laplace equation domain plus boundary conditions used to model the influence of the $MnO_2$ conductivity on the current density distribution. $r_{cut}$ denotes the cutting radius of the cut particle and $r_0$ the domain radius, see equation (5).

Figure 5. Diffusion domain plus boundary conditions used to simulate the influence of sulphuric acid surface concentrations on the current density distribution. $r_{cut}$ is the cutting radius and $r_0$ the domain radius.

Figure 6. Diffusion domain showing the boundary conditions of the hypothetical two-step two-material mechanism. $r_{cut}$ is the cutting radius and $r_0$ the domain radius.

Figure 7. a) Simulated TPBL plotted against $1/r$ with varying volume percentages of particles. b) Simulated TPBL versus TPBL calculated with equation (30).

Figure 8. Mechanism 1: Influence on current density of a) three different conductivities of the catalyst particles at $E - E^o = 0.6$ V for different particle sizes b) varying electrode potentials at $\sigma_2 = 1$ S/m for different particle sizes c) Radial current distribution for a 10 μm radius $MnO_2$ particle in a 20 μm domain for varying electrode potentials. d) Experimental data - effect of particle radius for varying electrode potentials vs. SHE, from ref. [17]. $r_p$ is the particle radius and $r_{rad}$ the radial coordinate.

Figure 9. Mechanism 2: Simulated transport of protons at an electrode with 25 vol% of $MnO_2$ particles. a) Varying electrode potential, $E - E^o$ and particle size, $r_p$. b) Radial current density distribution on a 10 μm $MnO_2$ particle with varying electrode potential. $r_{rad}$ is the radial coordinate.

Figure 10. Mechanism 3: Simulation results with respect to electrode potentials $E - E^o$ with varying decomposition rate constant $k_2$. a) Tafel plots. b) Tafel slope. c) Steady state bulk concentrations of $H_2O_2$. Equilibrium concentrations calculated by the Nernst equation are indicated by the dashed-dotted line.

Figure 11. Mechanism 3: a) Particle size ($r_p$) influence as a function of potential, $E - E^o$. b) Experimental results from [17]. c) Normalised radial current distribution on the lead matrix component at $E - E^o = -10$ mV. d) Corresponding surface concentrations of $H_2O_2$. $r_{rad}$ is the radial coordinate.

Figure 12. Mechanism 3: a) Current density versus total surface coverage $\Theta$, for different potentials, $E-E^o$ b) Plot of $J_{max}$ versus $\Theta_{max}$. Particles size is 10 μm.

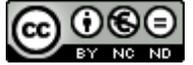

## *Figures*

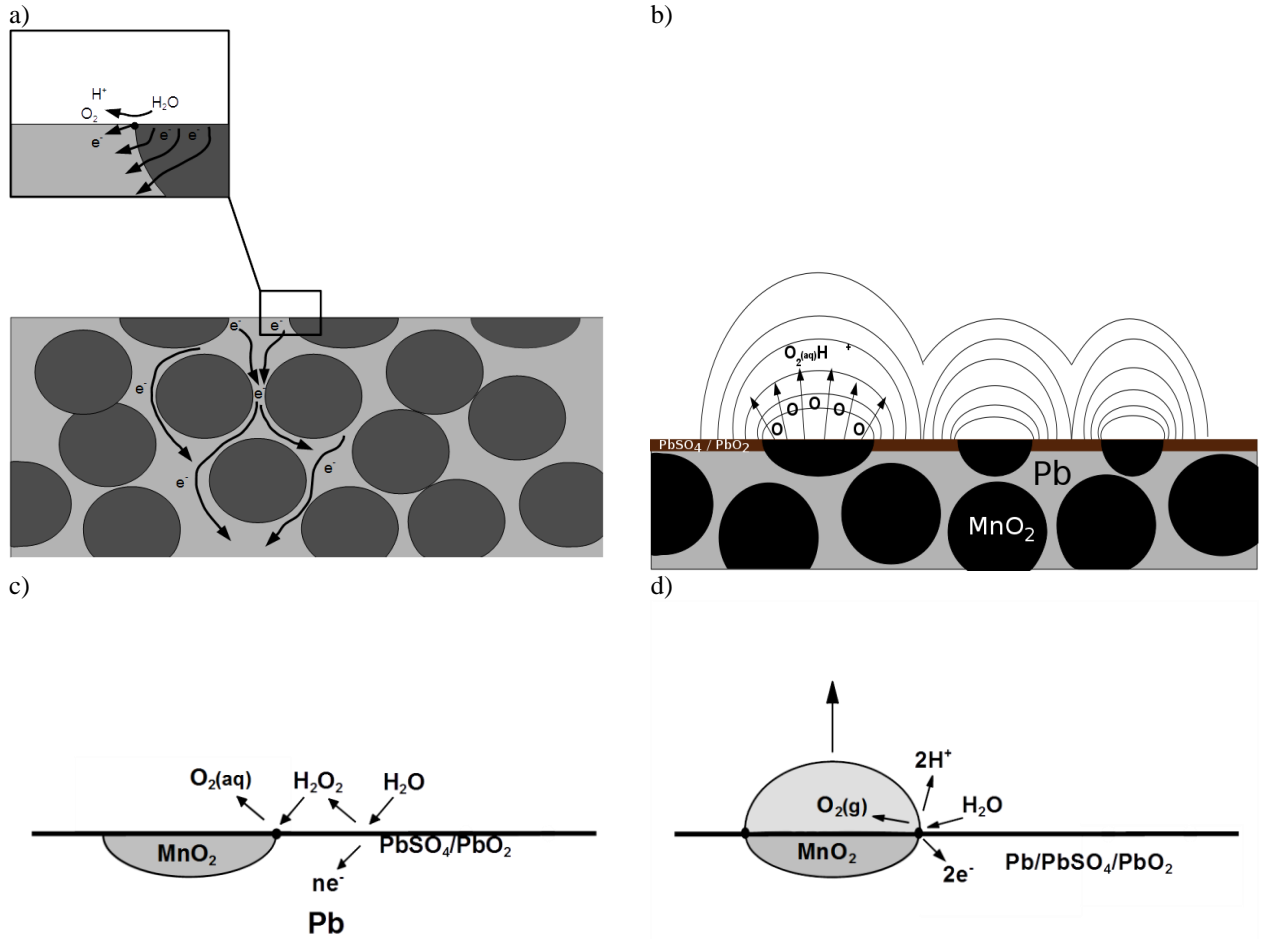

Figure 1

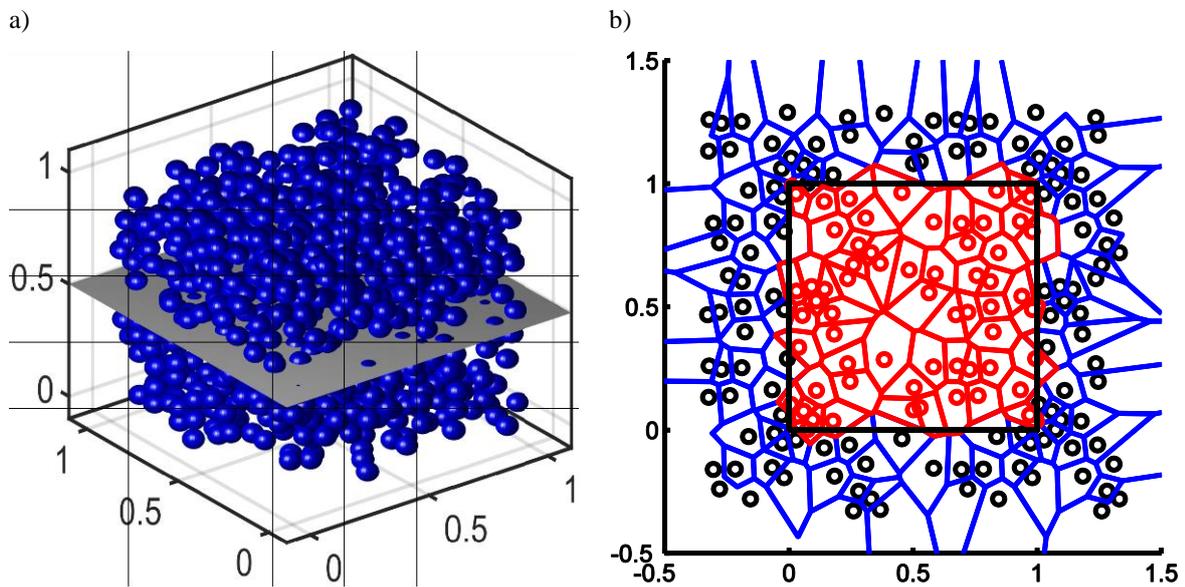

Figure 2

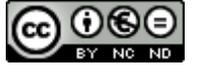

a)

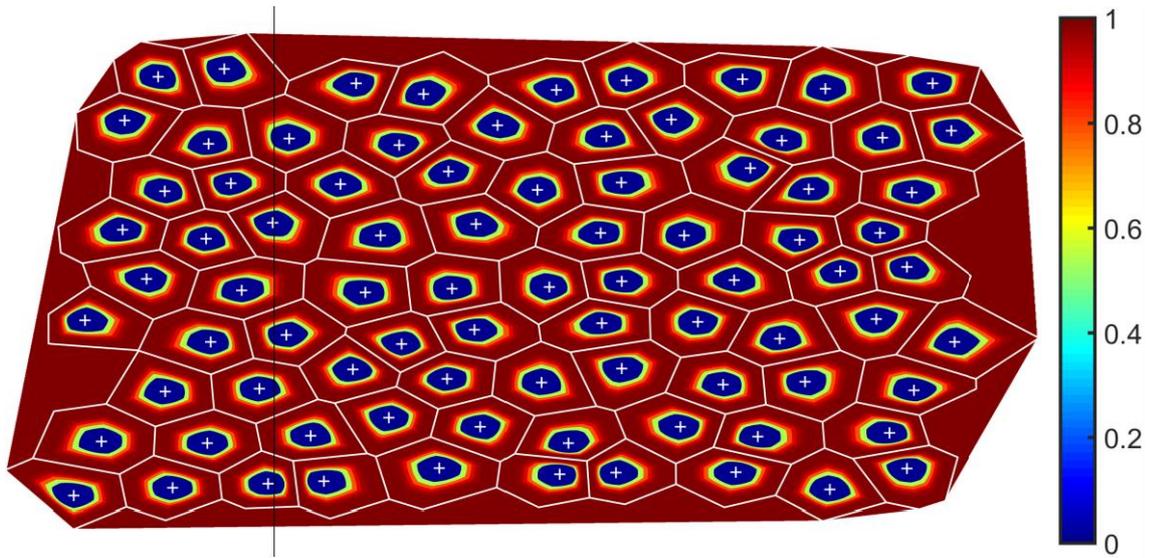

b)

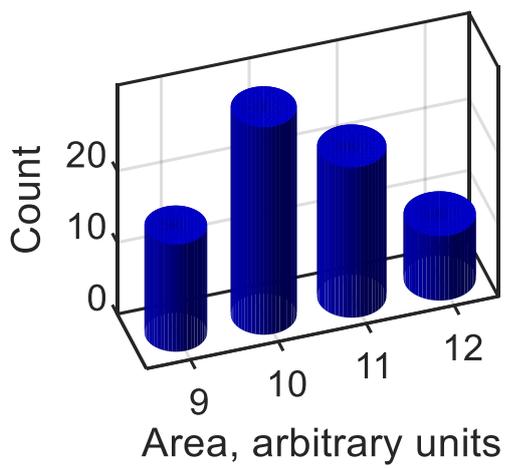

c)

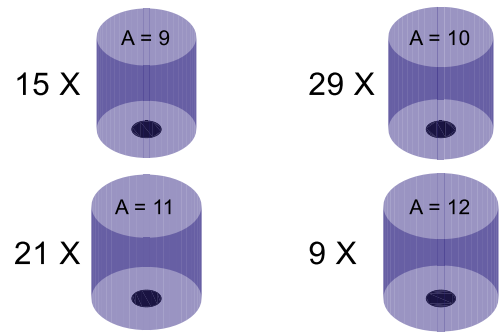

Figure 3

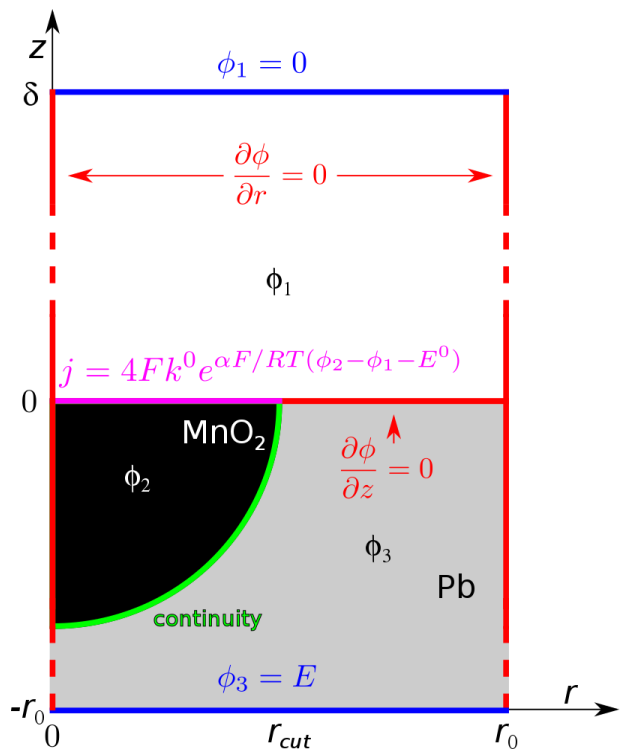

Figure 4

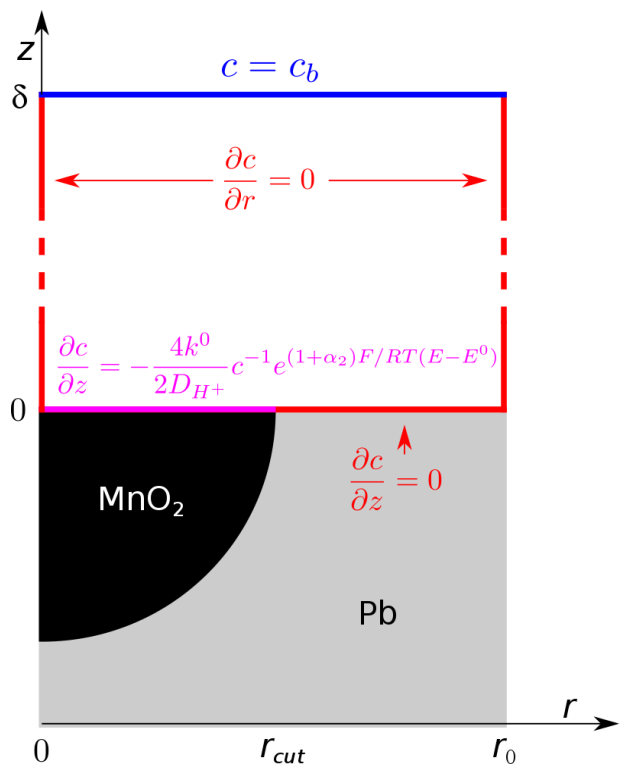

Figure 5

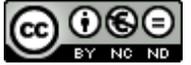

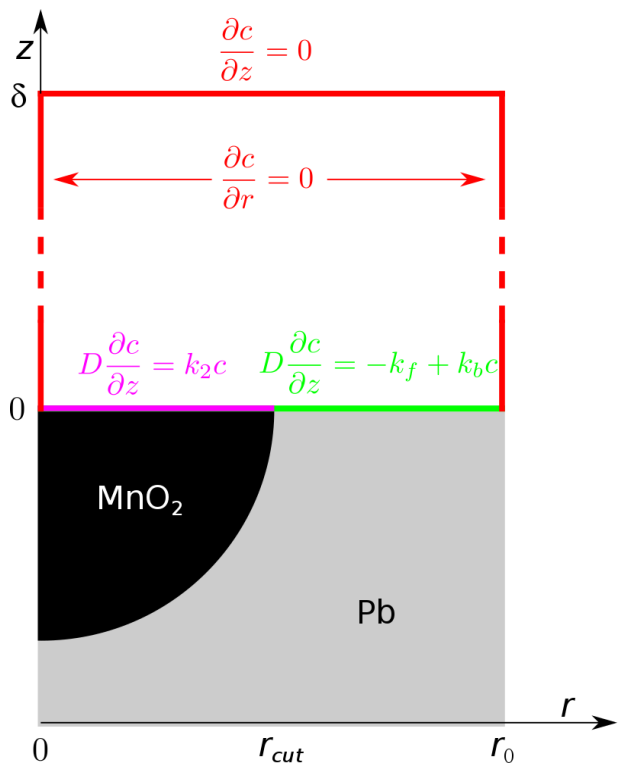

Figure 6

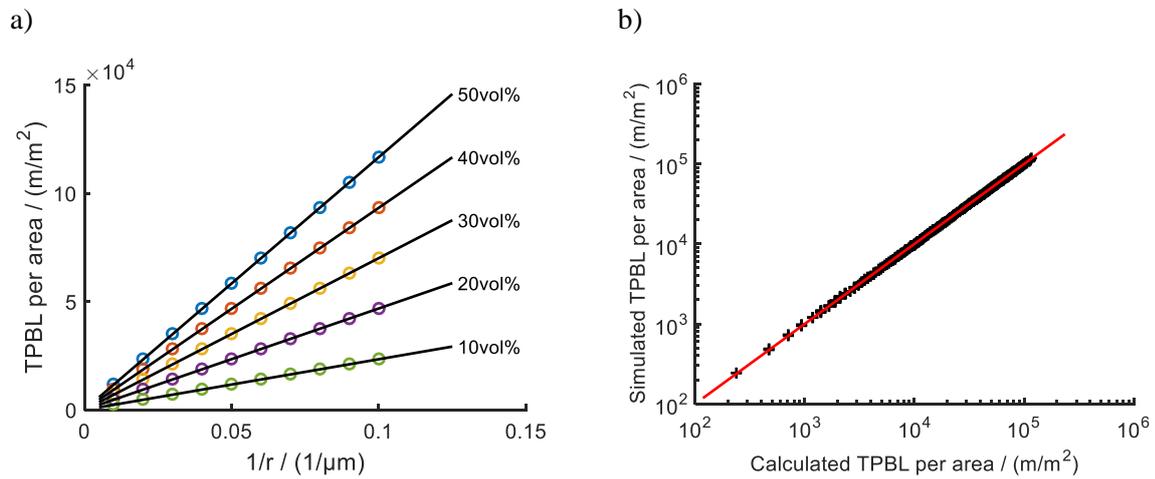

a)

b)

Figure 7

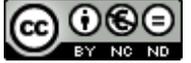

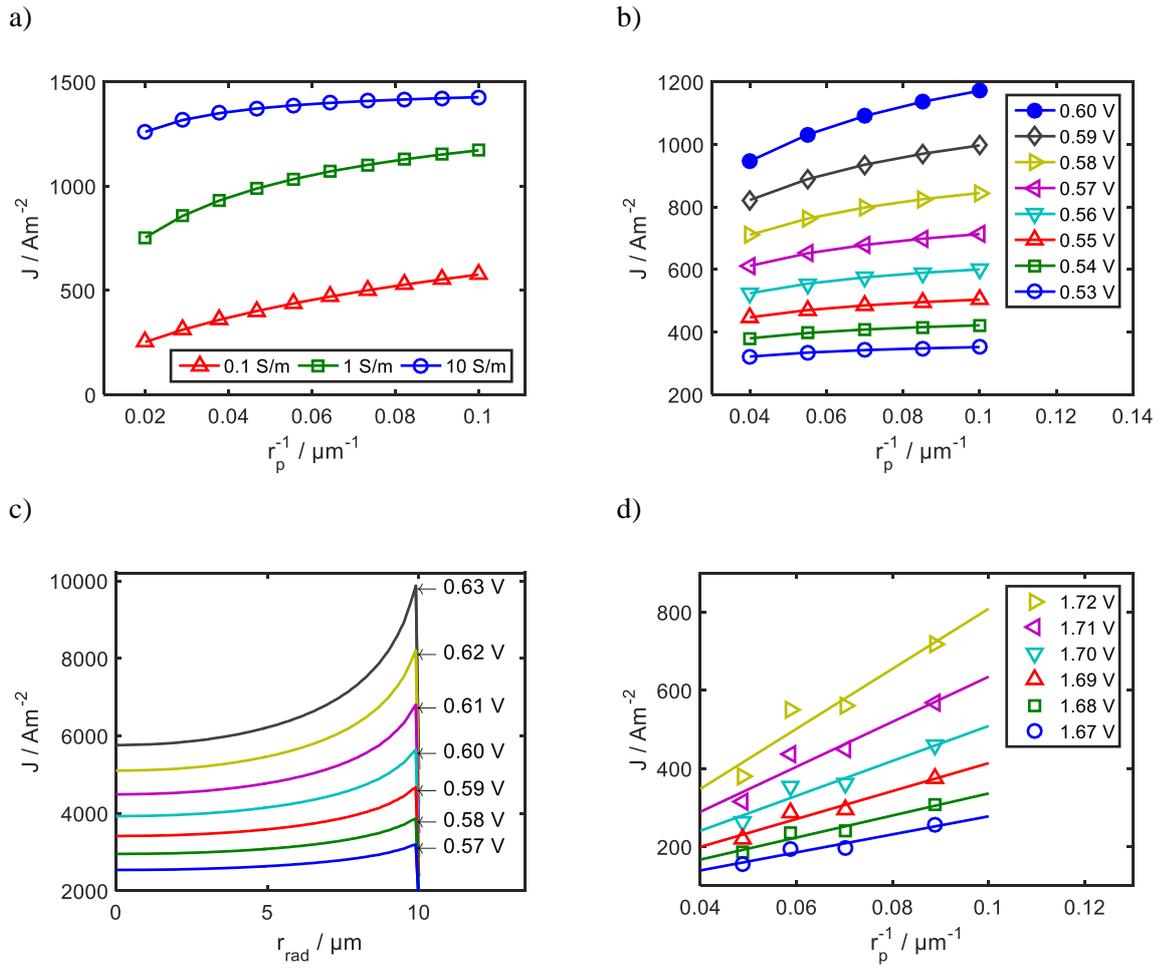

Figure 8

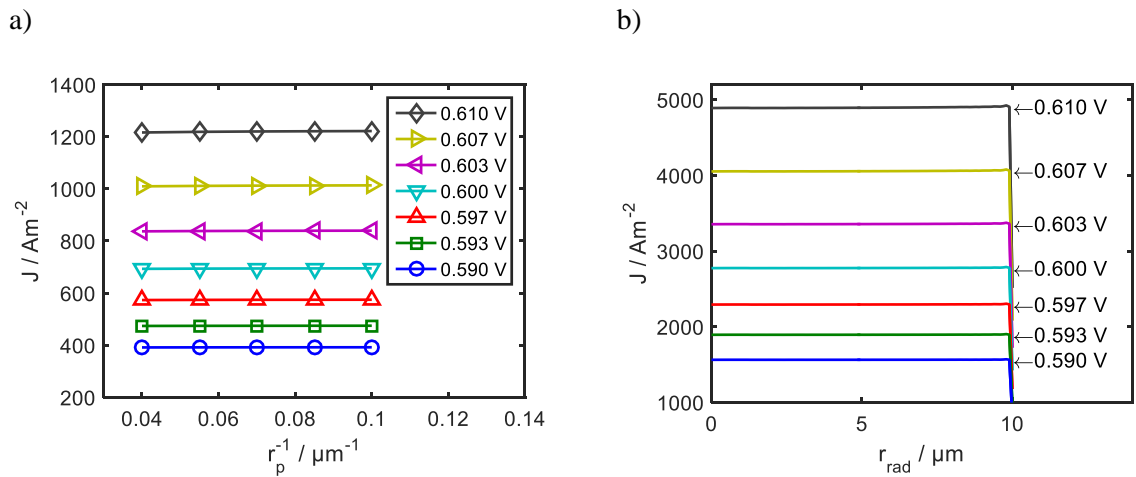

Figure 9

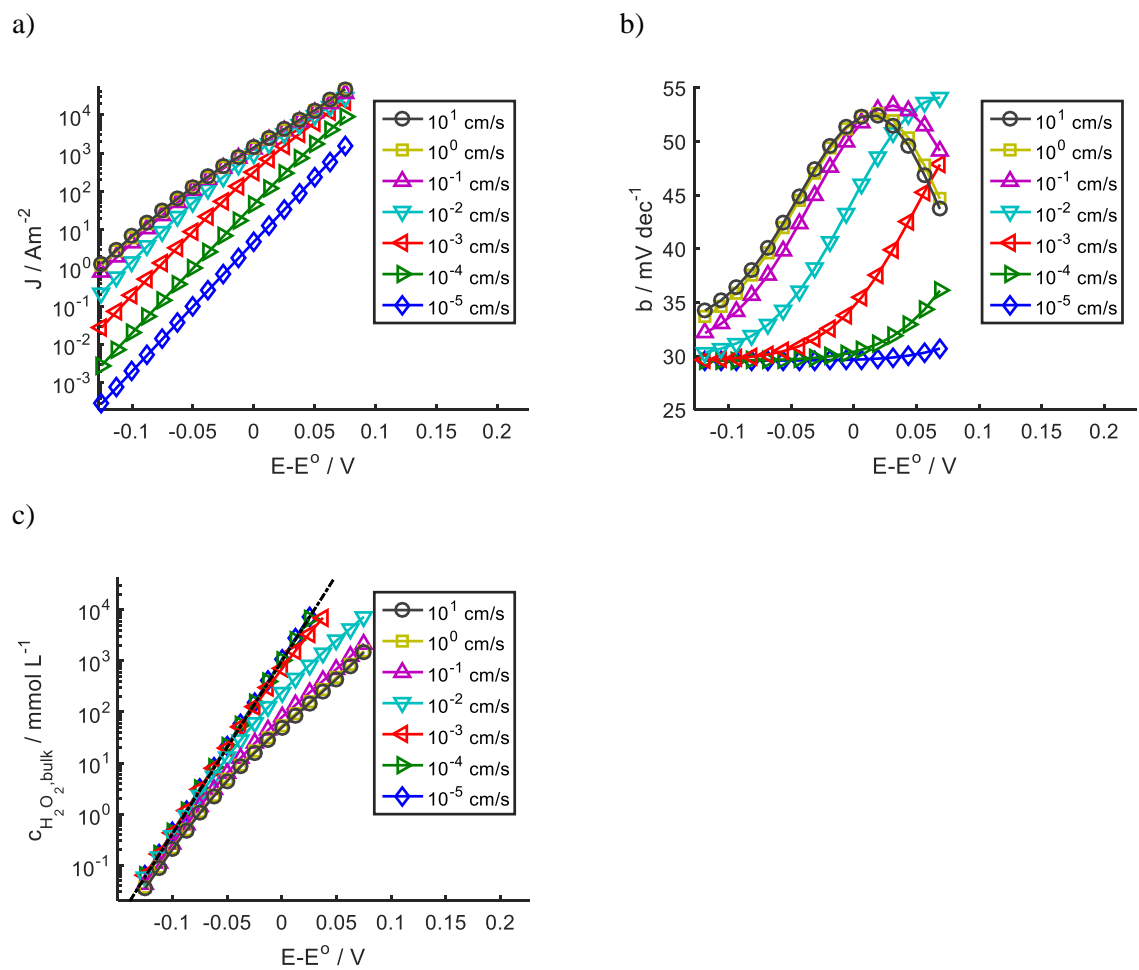

Figure 10

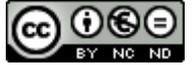

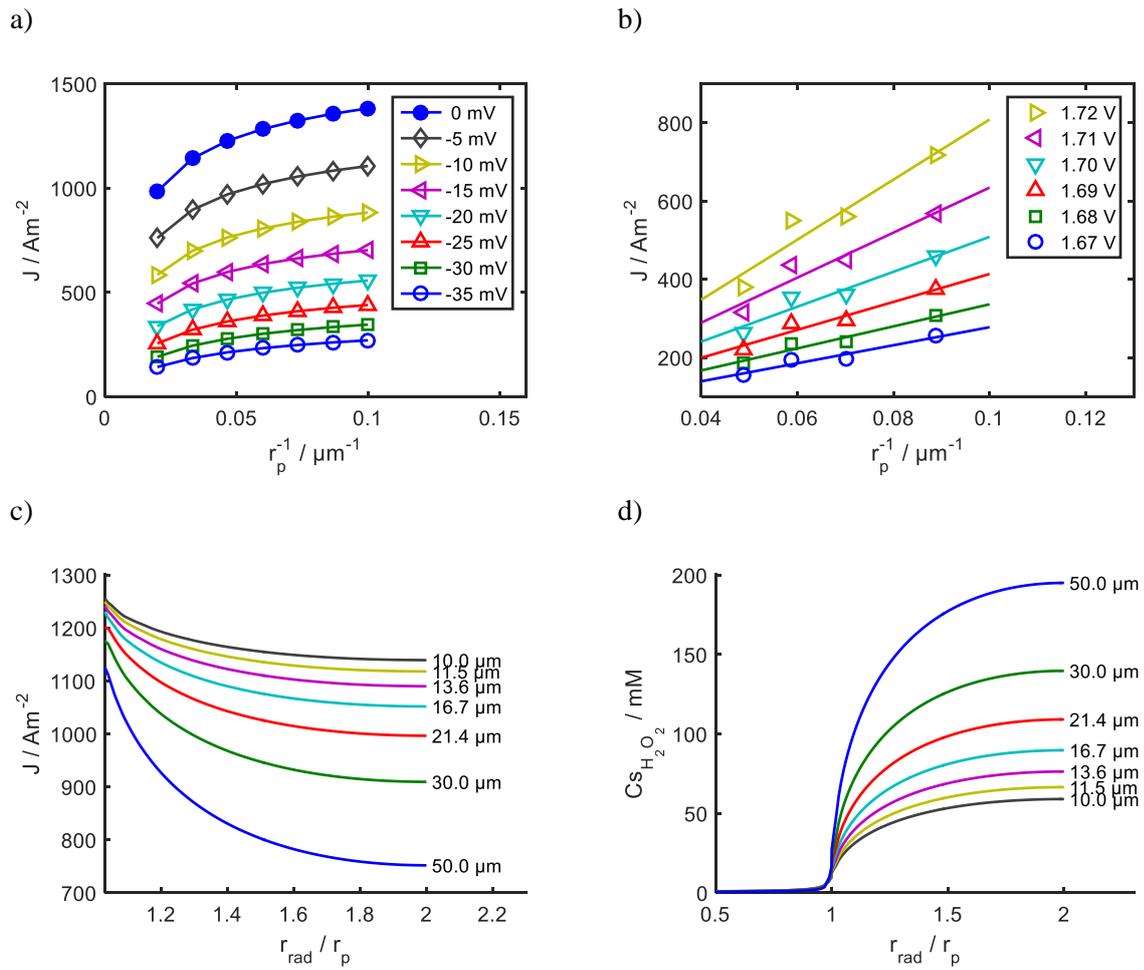

Figure 11

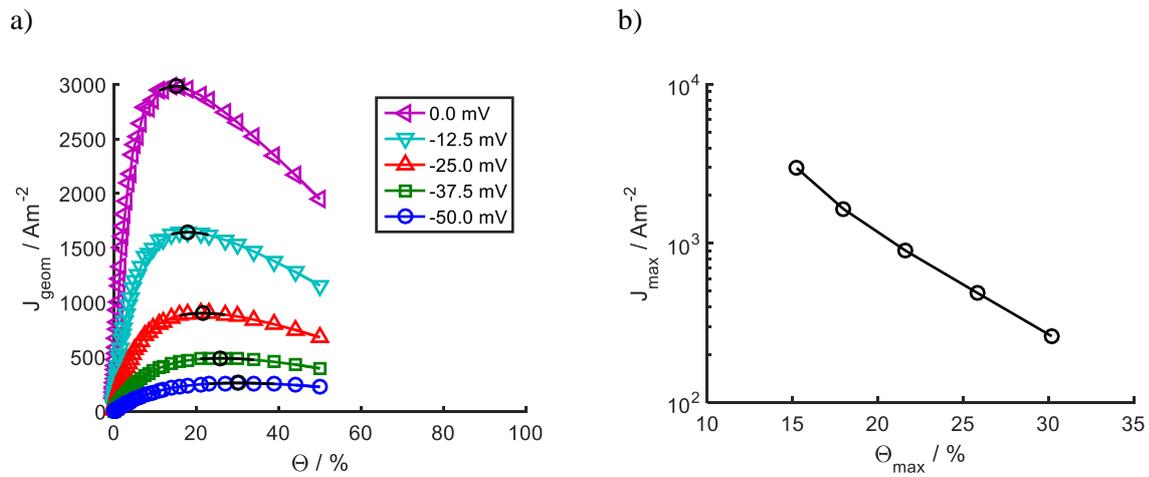

Figure 12

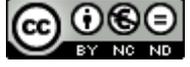

## Simulation of electrochemical processes during oxygen evolution on Pb-MnO₂ composite electrodes, Supplementary material

### 1. Triple phase boundary length model

The triple phase boundary length is derived from particles of the radius $r_p$ that are distributed in a cuboid. The cuboid is cut at the height $h$ that is chosen randomly: $r_p \geq h \geq -r_p$.

The average cut radius and cut area are calculated with equations (1) and (2).

$$\bar{r}_{cut} = \frac{1}{2r_p} \int_{-r_p}^{r_p} \sqrt{r_p{}^2 - h^2} \, dh = \frac{\pi}{4} r_p = \frac{1}{N} \sum_i r_{cut,i} \tag{1}$$

$$\bar{A}_{cut} = \frac{\pi}{2r_p} \int_{-r_p}^{r_p} \left( r_p{}^2 - h^2 \right) dh = \frac{2}{3} \pi r_p{}^2 = \frac{1}{N} \sum_i A_{cut,i} \tag{2}$$

The global surface coverage, $\Theta$ that is equal to the total surface fraction is calculated as

$$\Theta = \frac{\sum A_{cut,i}}{A_{geom}} = \frac{\bar{A}_{cut} N}{A_{geom}} \tag{3}$$

When $\bar{A}_{cut}$ in equation (3) is replaced by $\frac{2}{3}\pi r^2$, the number of cut particles, $N$ can be calculated as a function of the particle radius, $r_p$ the global surface coverage, $\Theta$ and the geometric surface area, resulting in equation (4).

$$N = \frac{\Theta \cdot A_{geom}}{\bar{A}_{cut}} = \frac{3}{2\pi} \frac{\Theta A_{geom}}{r_p{}^2} \tag{4}$$

The total length, $L$ of the triple phase boundaries is calculated from equation (5).

$$L = \sum_i l_i = 2\pi \sum_i r_{cut,i} = 2\pi N \bar{r}_{cut} = \frac{3}{4} \pi \frac{\Theta A_{geom}}{r_p} \tag{5}$$

### 1.1. Volume fraction and global surface coverage of composites

It can be shown that the global surface coverage, $\Theta$ is the same as the volume fraction, *vol%*. Consider a volume element $V$ with the thickness $t = 2r$, enclosing the surface of a composite electrode situated at $h = 0$ (Figure 1a). The number of particles, $M$, in that volume is calculated in a similar manner as above. The average volume of particles in that volume, $\bar{V}_p$ and the number of particles in volume $V$ are:

$$\bar{V}_p = \frac{1}{4r_p} \int_{-2r_p}^{2r_p} V_p(h') \, dh' = \frac{1}{2r_p} \int_0^{2r_p} \frac{(2r - h)^2}{3} (r + h) \, dh = \frac{2}{3}\pi r_p{}^3 \tag{6}$$

$$M = \frac{vol\% \cdot V}{\bar{V}_p} = \frac{3}{2\pi} \frac{vol\% \cdot A_{geom} 2r_p}{r_p{}^3} = \frac{3}{\pi} \frac{vol\% \cdot A_{geom}}{r_p{}^2} \tag{7}$$

In equation (6), $V_p(h)$ describes the intersection volume of a sphere with volume $V$, where the sphere resides at height $h$. The intersecting volume is described by the volume of a spherical cap, shown in Figure 1b, with the height $h' = |2r - h|$. Since the cuts are the same above ($h > 0$) and below the surface ($h < 0$), the average volume $\bar{V}_p$ in the interval [-2r,2r] is the same as the mean of $V_p(h)$ in the interval [0,2r].

Particles that stick into that volume are situated at heights, $2r_p \geq h \geq -2r_p$, (spheres 1-5 in Figure 1a). But only the spheres closer to the surface (2-4) are cutting the composite surface; for $r_p \geq h \geq -r_p$. Since the particles are distributed uniformly in the volume the number of cut particles is half of those being in the volume, $N = \frac{1}{2} M$. Combining equation (4) with the half of equation (7) yields:

$$N = \frac{3}{2\pi} \cdot \frac{\Theta \cdot A_{geom}}{r_p^2} = \frac{1}{2}M = \frac{3}{2\pi} \cdot \frac{vol\% \cdot A_{geom}}{r_p^2} \tag{8}$$

Since all terms in equation (8) are equal, it can be concluded that the volume fraction *vol%* must be equal to the global surface coverage $\Theta$.

## 2. Derivation of the oxygen evolution kinetics

The Tafel slope and the apparent reaction order of proton in OER mechanisms are usually calculated involving Langmuir isotherms, considering varying surface coverage of intermediates. Other steps than the rate determining step are assumed to be in equilibrium [1].

The electrochemical oxide path is given by the reactions (9)-(11) [1,3].

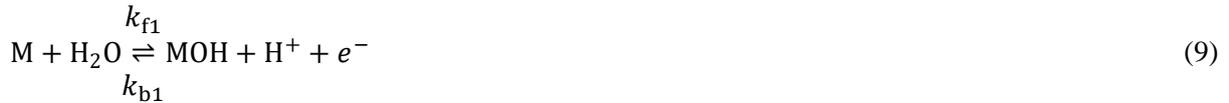

$$M + H_2O \underset{k_{b1}}{\overset{k_{f1}}{\rightleftharpoons}} MOH + H^+ + e^- \tag{9}$$

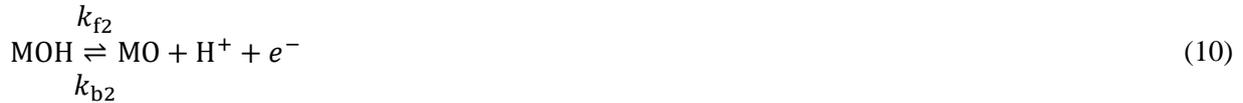

$$MOH \underset{k_{b2}}{\overset{k_{f2}}{\rightleftharpoons}} MO + H^+ + e^- \tag{10}$$

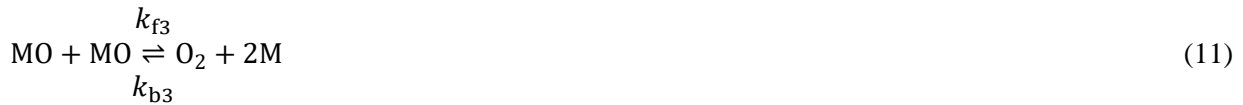

$$MO + MO \underset{k_{b3}}{\overset{k_{f3}}{\rightleftharpoons}} O_2 + 2M \tag{11}$$

Surface coverages of the adsorbates MO and MOH (M is a free site) are given by the Langmuir isotherm. Rate constants $k_{fi}$ and $k_{bi}$ in equations (9)-(11) depend on the electrode potential according to equations (12) and (13).

$$k_{fi}(E) = k_i^{0'} \cdot \exp\left(\frac{\alpha_i F}{RT}\left(E - E_i^0\right)\right) \tag{12}$$

$$k_{bi}(E) = k_i^{0'} \cdot \exp\left(-\frac{(1 - \alpha_i)F}{RT}\left(E - E_i^0\right)\right) \tag{13}$$

$\alpha_i$ is the charge transfer coefficient and $k_i^{0'}$ the standard rate constant of the respective step.

Similarly, the equilibrium constant $K_i'$ can be calculated with equation (14):

$$K_i'(E) = \frac{k_{fi}(E)}{k_{bi}(E)} = \exp\left(\frac{F}{RT}\left(E - E_i^0\right)\right) \tag{14}$$

It is possible to derive expressions in terms of the exchange current density of the respective steps, but great care should be taken in choosing their standard electrode potentials so that at equilibrium the

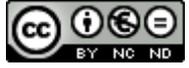

Nernst equation is fulfilled and no current flows. Instead, it is useful to rewrite equation (12) in terms of apparent rate constants $k_i^0$ that are related to the standard electrode potential of oxygen evolution:

$$k_{fi}(E) = k_i^{0\prime} \cdot \exp\left(\frac{\alpha_i F}{RT}(E - E_i^0)\right) = k_i^{0\prime} \cdot \exp\left(\frac{\alpha_i F}{RT}(E - E^0)\right) \cdot \exp\left(\frac{\alpha_i F}{RT}(E^0 - E_i^0)\right)$$
$$= k_i^0 \cdot \exp\left(\frac{\alpha_i F}{RT}(E - E^0)\right) \tag{15}$$

### 2.1. Mechanism 1, first step rate determining, secondary current distribution

The mechanism of Process 1 is based on the electrochemical oxide path with step 1 rate determining, where only one electron is transferred. Thus, following the apparent rate constant formulation above:

$$k_{f1}(E) = k_1^0 \cdot \exp\left(\frac{\alpha_1 F}{RT}(E - E^0)\right) \tag{16}$$

The backward reaction is neglected and the forward reaction determines the overall rate:

$$r = k_{f1}(E)(1 - \theta_{OH} - \theta_O)a_{H_2O} \tag{17}$$

The coverage by MOH and MO ($\theta_{OH}$ and $\theta_O$) depend on the equilibria of Steps 2 and 3 but they are assumed to be close to zero. Incorporating also the activity of water into the standard rate constant $k^0$, the current density is given by equation (18).

$$j = 4Fk^0 \exp\left(\frac{\alpha_1 F}{RT}(E - E^0)\right) \tag{18}$$

Note that the overall reaction has four electrons, but the rate determining step only one. Since reaction (9) was taken as irreversible it is not possible to express the current rate in terms of overpotentials and we are content to use the formulation of equation (18).

### 2.2. Mechanism 2, second step rate determining

With Step 2 rate determining, Steps 1 and 3 are taken to be in equilibrium. The total rate depends on the forward rate of Step 2, according to equation (19).

$$r = k_{f2}(E) \cdot \theta_{OH} = \theta_{OH} \cdot k_2^0 \cdot \exp\left(\frac{\alpha_2 F}{RT}(E - E^0)\right) \tag{19}$$

The coverage by MOH, $\theta_{OH}$, depends on the equilibria of Steps 1 and 3:

$$K_1'(E) = \frac{k_{f1}(E)}{k_{b1}(E)} = \exp\left(\frac{F}{RT}(E - E_1^0)\right) = \frac{a_H + \theta_{OH}}{a_{H_2O(1-\theta_{OH}-\theta_O)}} \approx \frac{a_H + \theta_{OH}}{a_{H_2O}} \tag{20}$$

$$K_3'(E) = \frac{k_{f3}(E)}{k_{b3}(E)} = \exp\left(\frac{F}{RT}(E - E_3^0)\right) = \frac{\frac{p_{O_2}}{p_0}(1-\theta_{OH}-\theta_O)^2}{\theta_O^2} \approx \frac{\frac{p_{O_2}}{p_0}}{\theta_O^2} \tag{21}$$

Assuming again that $1 - \theta_{OH} - \theta_O \approx 1$, $\theta_{OH}$ is given by equation (22), and the total rate by equation (23).

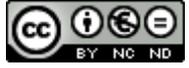

$$\theta_{OH} \approx \frac{K_1'(E) \cdot a_{H_2O}}{a_{H^+}} \tag{22}$$

$$r \approx k_{f2}(E) \cdot \frac{K_1'(E) \cdot a_{H_2O}}{a_{H^+}} = \frac{k^0}{c_{H^+}} e^{(1+\alpha_2) \cdot F/RT \cdot (E-E^o)} \tag{23}$$

The current density is obtained from equation (24). Assuming that the activity of water is one, the apparent standard rate constant, $k^0$, includes the rate constant $k_{f2}(E)$, the equilibrium constant $K_1'(E)$, and the activity coefficient of the proton.

$$j = 4Fk^0 c_{H^+}{}^{-1} \cdot e^{(1+\alpha_2) \cdot F/RT \cdot (E-E^o)} \tag{24}$$

## 2.3. Mechanism 1, potential applied at the MnO$_2$ particle surface

The equations to be solved are given in the main text, equations (7)-(13). The potentials can be explained with the reference to a virtual NHE electrode residing at $z = \delta$. Now, the total domain voltage is described by the following equation:

$$E = E_{boundary} + iR \tag{25}$$

$E_{boundary}$ describes the potential available to drive the electrochemical reaction and the $iR$ drop is the sum of the potential differences within the electrode, $E - \phi_2$, and the electrolyte, $\phi_1$:

$$iR = E - \phi_2 + \phi_1 \tag{26}$$

$$E_{boundary} = E - iR = \phi_2 - \phi_1 \tag{27}$$

This is the applied potential in the boundary value problem. The potential difference $\phi_2 - \phi_1$ thus refers to the applied potential corrected for the $iR$. The current density in equation (28) can then be determined from equation (27).

$$j = 4Fk^0 \exp\left(\frac{\alpha F}{RT}\left(E_{boundary} - E^0\right)\right) = 4Fk^0 \exp\left(\frac{\alpha F}{RT}\left(\phi_2 - \phi_1 - E^0\right)\right) \tag{28}$$

### Solving for the potential drop in a binary system

The Nernst-Planck equations in 1D can be solved in closed form, as follows:

$$\begin{cases} -\dfrac{\mathbf{N_+}}{D_+} = \dfrac{dc_+}{dz} + z_+ fc_+ \dfrac{d\phi}{dz} = -\dfrac{\mathbf{j}}{D_+ F} \\ -\dfrac{\mathbf{N_-}}{D_-} = \dfrac{dc_-}{dz} + z_- fc_- \dfrac{d\phi}{dz} = 0 \end{cases} \tag{29}$$

because only the cation is electroactive and carries all the current density $\mathbf{j}$ at steady-state. Therefore, the flux of the anion, $\mathbf{N_-}$, is zero everywhere, and the potential gradient is obtained from the above equation ($c_+ = c_- = c_1$) as [4]:

$$f \, d\phi = d\ln c_1 \quad \Rightarrow \quad \Delta\phi = \phi_1(z=0) = \frac{RT}{F} \ln\left(\frac{c_1^{surface}}{c_1^{bulk}}\right) \tag{30}$$

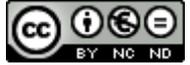

since we defined that $\phi_1(z = \delta) = 0$. The gist of this equation is that it includes both the diffusion potential and the iR drop in the solution, in our case [4]:

$$\nabla\phi = -\frac{RT}{F}\left(t_+ - t_-\right)\nabla\ln c - \frac{\mathbf{j}}{\kappa} \tag{31}$$

The concentration gradient is linear, which is seen summing

$$-\left(\frac{\mathbf{N}_+}{D_+} + \frac{\mathbf{N}_-}{D_-}\right) = -\frac{\mathbf{j}}{D_+F} = 2\frac{dc_1}{dz} \tag{32}$$

Integrating equation (32) from 0 to $\delta$ gives finally (no vector notation):

$$\phi_1(z = 0) = \frac{RT}{F}\ln\left(1 + \frac{j\delta}{2FD_+c_1^{\text{bulk}}}\right) \tag{33}$$

In our simulations, the value of the potential $\phi_1$ is less than 1 mV at worst.

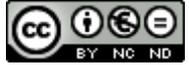

## *Figures*

a)
b)

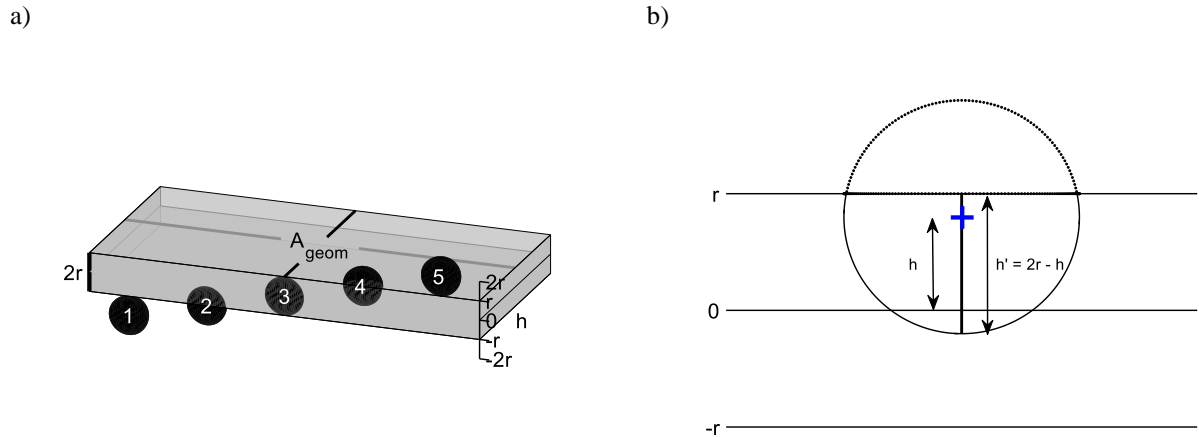

Figure 1. Particles within a layer around the surface at $h = 0$.  a) Dimensions of Volume $V$, Particles within $[-2r_p, 2r_p]$, spheres 1-5, contribute to the Volume, but only those within $[-r_p, r_p]$ are cut at $h = 0$ (spheres 2-4). b) Volume of a particle contributing to Volume $V$ (cross-section of a sphere cap). Volume situated lower than $h = r$ is included.

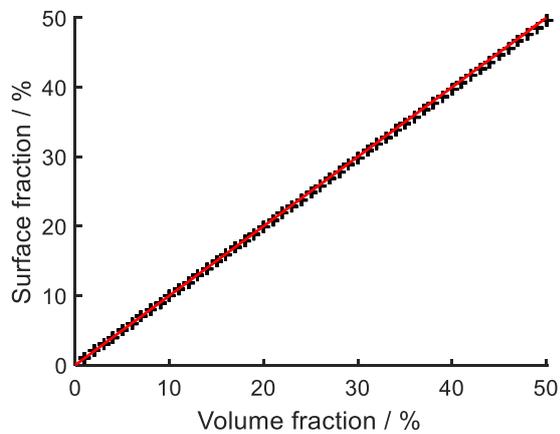

Figure 2. Surface fraction calculated from the simulation plotted against used volume fraction.